\documentclass[twocolumn,prb,10pt,aps,longbibliography,superscriptaddress]{revtex4-2}

\usepackage{graphicx}
\usepackage{grffile}
\usepackage{amsmath}
\usepackage{amssymb}
\usepackage{amsfonts}
\usepackage{dsfont}
\usepackage{dcolumn}
\usepackage{bbm}
\usepackage{bm}
\usepackage[caption=false]{subfig}

\usepackage{ulem}
\usepackage{mathtools}
\usepackage{braket}
\usepackage{placeins}
\usepackage{physics}
\usepackage{siunitx}
\usepackage[unicode]{hyperref}
\usepackage[nameinlink]{cleveref}
\Crefname{section}{Sec.}{Secs.}
\Crefname{equation}{Eq.}{Eqs.}
\Crefname{figure}{Fig.}{Figs.}
\Crefname{tabular}{Tab.}{Tabs.}

\usepackage{bbold} 
\usepackage{nicefrac}
\usepackage{tikz}
\usetikzlibrary{arrows.meta}
\usepackage{lipsum}  
\usepackage{xcolor}
\usepackage[version=4]{mhchem}

\usepackage[makeroom]{cancel}

\usepackage{standalone}

\definecolor{nat_green}{HTML}{43B02A}

\definecolor{tdo_green}{HTML}{83B818}
\definecolor{tdo_darkgreen}{HTML}{839A00}

\definecolor{rw_red}{HTML}{C8102E}
\definecolor{rw_darkred}{HTML}{971B2F}

\definecolor{med_blue}{HTML}{00A3E0}
\definecolor{med_darkblue}{HTML}{0061A0}

\definecolor{tdo_orange}{HTML}{D98207}
\definecolor{tdo_darkorange}{HTML}{CA7406}

\renewcommand\vec{\mathbf}

\usepackage{xstring}
\newcommand{\refcite}[1]{%
\begingroup
\def\tempx{0}%
  \StrCount{#1}{,}[\tempx]%
  \ifnum\tempx > 0 
  Refs.~%
  \else
  Ref.~%
  \fi
\endgroup
\cite{#1}%
}

\newcommand{\bes}{\begin{subequations}}
\newcommand{\ees}{\end{subequations}}
\newcommand{\be}{\begin{equation}}
\newcommand{\ee}{\end{equation}}

\newcommand{\dsc}{d_\text{sc}}
\newcommand{\jj}{$J_1$-$J_2$\,}

\NewDocumentCommand\alp{ s g }{ \ensuremath{ \IfBooleanTF#1
{ \alpha^{\dagger}_{#2}    } { \alpha^{\phantom{\dagger}}_{#2} } } }

\NewDocumentCommand\bet{ s g }{ \ensuremath{ \IfBooleanTF#1
{ \beta^{\dagger}_{#2}    }{ \beta^{\phantom{\dagger}}_{#2} } }} 

\NewDocumentCommand\alpeff{ s g }{ \ensuremath{ \IfBooleanTF#1
{ \alpha^{\dagger}_{#2}    } { \alpha^{\phantom{\dagger}}_{#2} } } }

\NewDocumentCommand\beteff{ s g }{ \ensuremath{ \IfBooleanTF#1
{ \tilde\beta^{\dagger}_{#2}    }{ \tilde\beta^{\phantom{\dagger}}_{#2} } }} 

\NewDocumentCommand\idel{}{\ensuremath{i+\delta}}

\NewDocumentCommand\abos{ s g }{ \ensuremath{ \IfBooleanTF#1
{ a^{\dagger}_{#2}    } { a^{\phantom{\dagger}}_{#2} } } }

\NewDocumentCommand\bbos{ s g }{ \ensuremath{ \IfBooleanTF#1
{ b^{\dagger}_{#2}    } { b^{\phantom{\dagger}}_{#2} } } }

\DeclareMathOperator{\sign}{sign}

\NewDocumentCommand\vect{g} {\ensuremath{  \bm{#1}  } }
\NewDocumentCommand\mat{g} {\ensuremath{  \text{\textbf{#1}}  } }

\NewDocumentCommand\cexpval{g}{\ensuremath{  \Braket{#1}}_0 }
\NewDocumentCommand\norord{g}{\ensuremath{:\!{#1}\!:}}

\NewDocumentCommand\NP{g}{\ensuremath{ N_{\text{p}} } }
\NewDocumentCommand\NAP{g}{\ensuremath{ N_{\text{ap}} } }

\newcommand{\eee}{\ensuremath{\text{e}}}
\newcommand{\ii}{\ensuremath{\text{i}}}

\newcommand{\disp}{\ensuremath{\omega}}

\NewDocumentCommand\boundstate{g}{\ensuremath{ \tau(#1) } }

\begin{document}

    \title{Quantum melting of long-range ordered quantum antiferromagnets investigated by momentum-space continuous similarity transformations}

    \author{Dag-Bj\"orn Hering}
    \email{dag.hering@tu-dortmund.de}
    \affiliation{Condensed Matter Theory, 
    Technische Universit\"{a}t Dortmund, Otto-Hahn-Stra\ss{}e 4, 44221 Dortmund, Germany}

    \author{Matthias R.\ Walther}
    \email{matthias.walther@fau.de}
    \affiliation{Department of Physics, Friedrich-Alexander-Universit\"{a}t Erlangen-N\"urnberg (FAU), Staudtstra\ss{}e 7, 91058 Erlangen, Germany}

    \author{Kai P.\ Schmidt }
    \email{kai.phillip.schmidt@fau.de}
    \affiliation{Department of Physics, Friedrich-Alexander-Universit\"{a}t Erlangen-N\"urnberg (FAU), Staudtstra\ss{}e 7, 91058 Erlangen, Germany}

    \author{G\"otz S.\ Uhrig}
    \email{goetz.uhrig@tu-dortmund.de}
    \affiliation{Condensed Matter Theory, 
    Technische Universit\"{a}t Dortmund, Otto-Hahn-Stra\ss{}e 4, 44221 Dortmund, Germany}

    \date{\textrm{\today}}

    \begin{abstract}
    We apply continuous similarity transformations (CSTs) to study the zero-temperature breakdown of long-range ordered quantum antiferromagnets.
    The CST flow equations are truncated in momentum space by the scaling dimension so that all operators with scaling dimension up to two are taken into account. 
    We determine the quantum phase transition out of the N\'eel-ordered phase in the unfrustrated square lattice Heisenberg bilayer as well as the quantum melting of the N\'eel-ordered and columnar phase in the highly frustrated \jj model on the square lattice.
    In all cases the CST is set up to isolate the ground state so that the stability of the flow equations, the ground-state energy, and the sublattice magnetization are used to explore the long-range ordered phases.
    We extract quantum-critical points which agree well with values in the literature.
    Further, we estimate the associated critical exponents $\alpha$ and $\beta$ which turns out to be a challenging task for the CST approach.
    \end{abstract}

    \maketitle

    \section{Introduction}
    \label{s:introduction}
		
    The collective behavior of interacting quantum matter has been an important topic in condensed matter physics in the last decades.
    Its study is key to the discovering and understanding of correlated many-body states with intriguing facets and the potential for new technological applications utilizing the quantum nature of these materials. 
    Examples are magnetic data storage \cite{chapp07}, spintronics, and platforms for quantum computing based on optical or solid state platforms \cite{baltz18,gomon17,bermu17,satzi21,bluvs24}.
    An important example is the discovery of high-$T_c$ cuprate superconductors \cite{bedno86}, 
    where the superconducting properties emerge in two-dimensional \ce{CuO2} layers \cite{manou91}. 
    Here, antiferromagnetic Heisenberg interactions between nearest-neighbor copper atoms are of central importance.  
	In extension, the antiferromagnetic Heisenberg bilayer is connected to high-temperature superconductors such as \ce{YBa2Cu3O6}$_{+x}$ and hence it is of fundamental interest to understand the influence of magnetic interactions in such two-dimensional systems \cite{Millis_1996}.

    In the undoped case, the low-energy physics is well described by an antiferromagnetic, nearest-neighbor spin-$\frac{1}{2}$ Heisenberg model
    The latter represents an interesting and challenging subject to study on its own. 
    On bipartite lattices, the zero-temperature ground state of the Heisenberg model displays long-range N\'eel order, where the SU(2) symmetry is spontaneously broken.
    Here the spins align antiferromagnetically with a preferred spin direction and sizeable quantum fluctuations are present in this two-dimensional system.
    The breaking of the continuous SU(2)-symmetry results in the presence of gapless Goldstone bosons, which are called magnons.
    While the low-energy properties of the magnons are well understood for quite some time, only recently a quantitative understanding of the high-energy part of the one-magnon dispersion such as the characteristic roton minimum has been achieved \cite{Powalski15, Powalski18, verre18b}.
    Apart from the one-magnon dispersion, also the full dynamical structure factor up to three-magnon continua has been calculated by continuous similarity transformations (CST) in momentum space \cite{Powalski15, Powalski18} yielding quantitative agreement with inelastic neutron scattering data of undoped cuprate materials.
    Technically, the CST is based on the non-Hermitian Dyson-Maleev transformation \cite{Dyson_1956,malee58b} 
		and the CST flow equations have been truncated by the scaling dimension of operators so that all quartic magnon-magnon interactions with scaling dimension up to two are taken into account.
    
    A next natural step is to investigate the breakdown of long-range ordered phases using the CST approach. 
    Keeping in mind that a truncation with scaling dimension up to one corresponds to the self-consistent mean-field solution, it is an interesting open question whether one can capture the quantum melting of long-range ordered phases quantitatively by CST when truncating on the level of quartic operators, which have scaling dimension two.

    In a first step in this direction, the CST has been applied successfully to the symmetry-enhancement transition in the antiferromagnetic, {spin-$\frac{1}{2}$} XXZ-model on the square lattice tuning from the spin-anisotropic Ising point to the spin-isotropic Heisenberg model \cite{Walther23}.
    Beyond the ground-state energy and the decay of two-magnon bound states, the CST turned out to  be able to quantitatively determine the algebraic behavior as well as its prefactor of the closing of the single-magnon gap.

    However, this gap closing is not a genuine quantum phase transitions since no symmetry is broken spontaneously and genuine critical fluctuations are absent.
    In fact, it is only the symmetry which is enhanced from $\mathbb{Z}_2$ to SU(2) when tuning the anisotropy from the Ising limit to the isotropic Heisenberg point.
    As a consequence, the associated exponents are of mean-field type.
    Quantitative agreement has been found between the CST and quantum Monte Carlo data (QMC) \cite{caci24}.
		
    In the present work we proceed further and apply the CST to two paradigmatic models where long-range antiferromagnetic order breaks down by quantum phase transitions. 
    Specifically, we investigate the unfrustrated, {spin-$\frac{1}{2}$} Heisenberg bilayer on the square lattice that features a well studied quantum phase transition in the O(3) universality class \cite{Campostrini_2002,Wang_06,hamer12,Lohoefer_15} separating the long-range N\'eel order from the featureless singlet phase at strong interlayer couplings.
    This system serves as a benchmark making quantitative comparisons to QMC studies \cite{Wang_06,Lohoefer_15} possible.
		
   Subsequently, we study the highly frustrated, {spin-$\frac{1}{2}$} \jj  model on the square lattice that features two magnetically ordered phases at small and large next-nearest neighbor interactions $J_2$. 
  The nature of the quantum critical points and the associated critical exponents as well as the quantum phases in the intermediate regime are still highly debated because the presence of geometric frustration in the exchange couplings leads to many competing ground states which are hard to capture quantitatively by any numerical technique.
   Here, we focus on the breakdown of the two magnetically ordered phases and compare our results with the ones from other approaches where available.

    The paper is structured as follows.
    In \Cref{s:model} we describe the relevant properties of the Heisenberg bilayer on a square lattice and of the \jj  model. 
    A description of the CST as well as of technical aspects is provided in \Cref{s:methods}; this section can be skipped by those who want to focus on the results.
    Results for the critical points and the critical exponents $\alpha$ and $\beta$ are provided in \Cref{s:results}.  We conclude our study in \Cref{s:conclusion}. 

    \section{Models}
    \label{s:model}
		In this section, we introduce the square lattice Heisenberg bilayer and the \jj  Heisenberg model on the square lattice, which we investigate in this article. In particular, we summarize the known quantum-critical properties for both systems.
		
        \subsection{Square lattice Heisenberg bilayer}
        \label{ss:hbbilayer}
\begin{figure}
                \centering
                \includegraphics[width=\columnwidth]{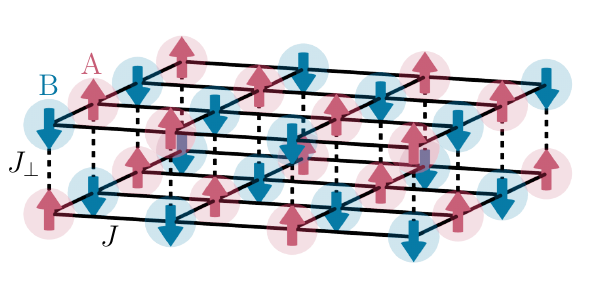}
                \caption{
                  Sketch of the magnetically ordered phase of the antiferromagnetic Heisenberg bilayer on the square lattice with black solid bonds representing the nearest-neighbor coupling $J$ and black dashed bonds representing the inter-layer coupling $J_{\perp}$.
                  Red and blue arrows illustrate the spin orientation on the different sublattices A and B in the ground state. 
                }
                \label{fig:HBBLattice}
            \end{figure}

            The Hamiltonian of the square lattice Heisenberg bilayer is given by
            \begin{align}
                \mathcal{H} = J \sum_{\alpha=1}^2\sum_{\langle i,j \rangle} \hat{S}_{i,\alpha} \hat{S}_{j,\alpha} + J_\perp \sum_i \hat{S}_{i,1} \hat{S}_{i,2}, \label{eqn:HBBilayer}
            \end{align}
            where $S_{i,\alpha}$ denotes the {spin-$\frac{1}{2}$} on site $i$ in layer $\alpha\in\{1,2\}$, $\langle i,j \rangle$ indicates a pair of nearest neighbors in each layer and $J$ the coupling between them, while $J_\perp$ is the nearest-neighbor inter-layer coupling.
            The  square lattice Heisenberg bilayer and the sublattices as used in our calculations are shown in \Cref{fig:HBBLattice}.
            For ${\lambda_\perp\coloneqq J_{\perp}/{J} = 0}$, the Hamiltonian reduces to two independent copies of the single-layer antiferromagnetic Heisenberg model on the square lattice studied in \refcite{Powalski15, Powalski18, Zheng_2005,Sandvik_2001} with broken SU(2) symmetry.  
			The system displays the long-range ordered N\'eel ground state with gapless Goldstone bosons, called magnons, as elementary excitations. 
            In the opposite limit, ${\lambda_\perp \gg 1}$, the ground state is a featureless gapped singlet state which is adiabatically connected to the limit of isolated dimers.

            Since this model is bipartite and unfrustrated, the quantum phase transition between these two phases has been studied reliably with QMC techniques and the critical point was found to be ${\lambda_{\perp,c} = \num{2.5220 \pm 0.0001}}$.
            The quantum critical properties are governed by the O(3) universality class \cite{Wang_06,Lohoefer_15,demid23} with critical exponents $\alpha=\num{0.1336 \pm 0.0015}$ and $\beta=\num{0.3689 \pm 0.003}$ \cite{Campostrini_2002}.
            This well-studied phase transition will serve as a benchmark for the CST study below.

            Previous spin-wave calculations strongly overestimate the location of the quantum critical point.
            Linear spin-wave theory yields ${\lambda_{\perp,c} \approx \num{13.6}}$ while self-consistent spin-wave theory and Schwinger-boson mean-field theory yield the improved values ${\lambda_{\perp,c} \approx \num{4.3}}$ \cite{Hida_1990} and ${\lambda_{\perp,c} \approx \num{4.5} }$ \cite{Millis_1993}, respectively. 
             In \refcite{Chubukov_1995}, it is argued that the main weakness of spin-wave theory consists in the fact that transversal and longitudinal fluctuations are not treated equally and an expansion around the non-magnetic phase is presented yielding ${\lambda_{\perp,c} \approx \num{2.73}}$.
          					
        \subsection{\texorpdfstring{\jj}{TEXT}  Heisenberg model on the square lattice}
        \label{ss:j1j2model}

        The Hamiltonian of the \jj  Heisenberg model on the square lattice is given by 
            \begin{align}
                \mathcal{H} = J_1 \sum_{\langle i,j \rangle>} \hat{S}_i \hat{S}_j + J_2 \sum_{\langle\langle i,j \rangle\rangle} \hat{S}_i \hat{S}_j , 
                \label{eqn:J1J2}
            \end{align}
            with antiferromagnetic couplings $J_1,J_2>0$ and ${\langle i,j \rangle}$ (${\langle\langle i,j \rangle\rangle}$) labels nearest neighbors (NN) [next nearest neighbors (NNN)]. 
            In contrast to the square lattice Heisenberg bilayer, this model is highly frustrated because of the simultaneous presence of the NNN antiferromagnetic coupling $J_2$ and the NN coupling $J_1$.
			Only if one of them vanishes, no geometric frustration is present.

            Despite its simplicity, the \jj  Heisenberg model represents an extremely challenging problem. 
            Its quantum phase diagram is expected to be rich with potentially exotic quantum spin liquid phases in the intermediate regime $J_1\approx J_2$ \cite{Sushkov01,richterSpin1J1J22010,jiangSpinLiquidGround2012,Doretto14,moritaQuantumSpinLiquid2015,Liu2022}.
            As a function of $\lambda_{12}=J_2/J_1$, the two limiting phases are the N\'eel phase at small values of $\lambda_{12}$ and the columnar phase at large $\lambda_{12}$.
            The quantum melting of these phases are the foci of the present work.
            The SU(2) symmetry-broken N\'eel phase is realized for $\lambda_{12} \lessapprox 0.4$ \cite{Sushkov01,richterSpin1J1J22010,jiangSpinLiquidGround2012,Doretto14,moritaQuantumSpinLiquid2015,Wang18,Liu2022} and is therefore adiabatically connected to the NN Heisenberg model on the square lattice, which we introduced already in \Cref{ss:hbbilayer}.
            For $\lambda_{12} \gtrapprox 0.6 $ \cite{jiangSpinLiquidGround2012,Wang18,richterSpin1J1J22010,Liu2022,moritaQuantumSpinLiquid2015}, the ground state of the system is a columnar ordered phase which is adiabatically connected to the limit $J_2 \gg J_1$.
            For $J_1=0$, the model reduces to two independent copies of the nearest-neighbor Heisenberg model where the CST approach yields quantitative results \cite{Powalski15, Powalski18, Walther23}.

			\begin{figure}
                \centering
                \includegraphics[width=\columnwidth]{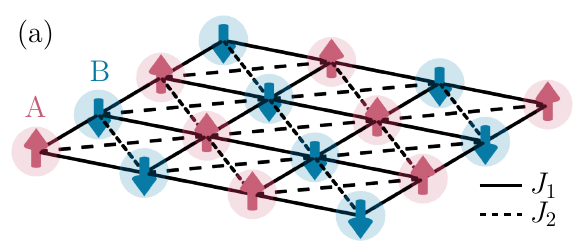}
				\includegraphics[width=\columnwidth]{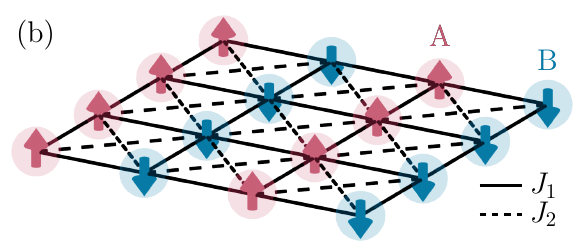}
            \caption{
                The panels show the two magnetically ordered phases of the antiferromagnetic \jj model with solid black bonds representing the nearest-neigbor interaction $J_1$ and the dashed black bonds representing the next-nearest neigbor interaction $J_2$.
                Red and blue arrows illustrate the spin orientation on the sublattices A and B in the ground state.  
                In panel (a), the N\'eel ordered phase is shown where $J_1\gg J_2$ and in panel (b) the columnar ordered phase  where $J_2\gg J_1$. 
            	}
                \label{fig:J1J2Lattice}
            \end{figure}

            For the intermediate regime $J_1\approx J_2$, the existence of disordered phases such as gapless or gapped quantum spin liquids or different valence-bond solids with columnar or plaquette order are discussed 
						\cite{Sushkov01,richterSpin1J1J22010,jiangSpinLiquidGround2012,Doretto14,moritaQuantumSpinLiquid2015,Liu2022}.
            There are indications that the phase transition from the columnar phase to a  disordered intermediate phase is weakly first order \cite{Sushkov01,moritaQuantumSpinLiquid2015,jiangSpinLiquidGround2012,richterSpin1J1J22010, Liu2022}.
            The corresponding numerical evidence is found using density matrix renormalization group (DMRG) \cite{jiangSpinLiquidGround2012,Wang18}, coupled cluster method (CCM) \cite{richterSpin1J1J22010}, projected entangled pair states (PEPS) \cite{Liu2022} and variational QMC \cite{moritaQuantumSpinLiquid2015}.
            For the phase transitions between the N\'eel state and a disordered state the situation is less clear.
            The authors of the \refcite{jiangSpinLiquidGround2012,richterSpin1J1J22010,Liu2022,moritaQuantumSpinLiquid2015} find indications of a continuous phase transition, but a contrary finding exists as well \cite{sirke06} and no conclusive results are available to date.

            Here, we apply the CST to both, the N\'eel and the columnar phase, and track the quantum melting of these long-range ordered gapless phases towards the intermediate regime.
            Linear spin wave theory (LSWT) in both ordered cases show a melting of the magnetic order indicated by the vanishing of the sublattice magnetizations \cite{Xu1990,uhrig09a}.
            For the N\'eel phase, the melting occurs at $\lambda_{12} \approx \num{0.38}$ and for  the columnar phase LSWT predicts $\lambda_{12} \approx 0.51$.
            In contrast, self-consistent mean-field spin wave theories find an intermediate region from $\lambda_{12} \approx \num{0.55}$ to $\lambda_{12} \approx \num{0.62}$, where both phases show non-zero magnetization so that they may exist. Hence, no indication about the nature of the intermediate phase can be deduced.

            The starting point for the N\'eel phase is shown in \Cref{fig:J1J2Lattice}(a) where the lattice is divided into two sublattices with alternating spins.
            The unit cell is the same as for the NN Heisenberg model, but one has to take the additional frustrating NNN couplings into account linking two spins of the same sublattice.
            In the limit of $\lambda_{12} \rightarrow \infty$, the NNN couplings dominate. 
            Eventually, two copies of the NN Heisenberg model arise on square lattices rotated by $45^\circ$ with lattice constant $\sqrt{2}a$. These two copies are coupled by the NN interaction.
            For $J_1=0$, two independent SU(2) symmetries are broken, but a residual NN coupling $J_1$ reduces the degeneracy and only two choices for the spin orientation of the second sublattice remain.
            As a reference state for out calculations, we choose the one shown in \Cref{fig:J1J2Lattice}(b).
            Note that the unit cell of the columnar phase is a $2a \times a$ rectangle instead of a $\sqrt{2}a \times \sqrt{2}a$ diamond.
 
    \section{Methods and Techniques} 
    \label{s:methods}
       
        \subsection{Continuous Similarity Transformation (CST)}
        \label{ss:cst}
        We use the CST in momentum space with a truncation in the scaling dimension as done in Refs.\ \cite{Powalski15, Powalski18, Walther23}.
        The main idea of flow equation based approaches \cite{Wegner_1994,Mielke_1998,Knetter_2000,Knetter_2003} is to transform a given initial Hamiltonian $\mathcal{H}_{0}$ in a continuous way into a basis where it is (more) diagonal.
        The Hamiltonian after the transformation is called the effective Hamiltonian $\mathcal{H}_{\text{eff}}$.
        An auxiliary flow parameter $\ell$ is introduced that parametrizes the transformation from $\mathcal{H}(\ell=0) = \mathcal{H}_0$ to  $\mathcal{H}(\ell=0) = \mathcal{H}_{\text{eff}}$ and the flow is given by the flow equation $\partial_{\ell} \mathcal{H}(\ell) = \left[ \eta(\ell), \mathcal{H}(\ell) \right]$ where $\eta(\ell)$ is the anti-hermitian infinitesimal generator of the transformation.
        Computing $\left[ \eta(\ell), \mathcal{H}(\ell) \right]$ generally yields infinitely many terms.
        In order to obtain a closed system of ordinary differential equations, a systematic and controlled truncation scheme must be utilized.
        The choice of a suitable generator and truncation scheme is central to flow-equation approaches.

        We use the same truncation scheme as in \refcite{Powalski15, Powalski18, Walther23} in terms of the scaling dimension of operators. Indeed, the scaling dimension $\dsc$ introduces a hierarchy among the operators where operators with lower scaling dimension are more relevant for the low-energy physics than operators with a larger scaling dimension.
        In \refcite{Powalski15, Powalski18} it was demonstrated that the CST with a truncation of operators with $\dsc > 2$ is able to reproduce the one-magnon dispersion and the dynamical structure factor of the Heisenberg antiferromagnet quantitatively.
        In \cite{Walther23} the same truncation was used to determine the critical exponents of the closing of the one-magnon gap quantitatively as well as the decay of two-magnon bound states into the two-magnon continuum.

        Truncating the flow for $\dsc>1$ is equivalent to mean-field theory where all phase transitions discussed in \Cref{s:model} are present, but with critical points inconsistent with state-of-the-art numerical studies \cite{Hida_1990,Xu1990}.
        Here we investigate how extending the truncation scheme considering all operators with $\dsc\le2$ improves the results of the mean-field approach and if critical points and critical exponents can be determined.

        In contrast to the former CST studies for ordered magnets \cite{Powalski15,Powalski18,Walther23,caci24} we here choose the $0n$-generator \cite{Fischer_2010} instead of the full quasiparticle-number conserving generator \cite{Mielke_1998, Knetter_2000}.
        The full quasiparticle-number conserving generator is given by $\eta_{ij} = \sign( q_{ii} - q_{jj}) h_{ij}$, where $q_{ii}$ are the eigenvalues of the quasi-particle counting operator $Q$ and $h_{ij}$ is a matrix element of $\mathcal{H}$ in an eigenbasis of $Q$.
        This generator allows one to disentangle all quasi-particle sectors so that $\mathcal{H}_{\text{eff}}$ acquires a block-diagonal form where each block contains elements that act only on a fixed number of quasi-particles without changing this number.
        However, this only succeeds if there are no energetic overlaps between different quasi-particle sectors.
        For example, the overlap of the ground-state energy sector (zero quasi-particles) with an energy in the one-quasi-particle (1QP) sector corresponds to the closing of the single-particle gap and hence a second-order phase transition.
        An overlap of sectors with higher quasi-particle number within a stable phase is also possible, for instance induced by binding effects, and will result in a divergent flow.  

        We encounter such energetic overlaps and associated divergences of the flow for all three long-range ordered magnetic phases under study when applying the quasi-particle generator. 
        Consequently, we use the $0n$-generator \cite{Fischer_2010} that only disentangles the ground state from all higher particle sectors. 
        Generically, this yields more robust flows since energetic overlaps in higher quasi-particle sectors do not matter .
        Still, the $0n$-generator allows us to study the phase transition out of the ordered phases by the stability of the flow and by the analysis ground-state properties such as its energy or the sublattice magnetization. This will be discussed in more detail in Sec.~\ref{ss:observables}.

        \subsection{Self-consistent mean-field approach}

        The starting point of the CST is the self-consistent mean-field solution in the thermodynamic limit.
        This approach has the crucial advantage that it captures the gapless Goldstone bosons for the 
				N\'eel and columnar phase, which is already challenging for an approach using flow equations represented in real space \cite{Schmidt_2006}.

        The reference state for the mean-field calculations is in all cases the classical ordered state with alternating spin up and down on the two sublattices. 
        Then we introduce bosonic degrees of freedom by the Dyson-Maleev transformation \cite{Dyson_1956,malee58b}. 
        This transformation is not unitary so that the resulting Hamiltonian is not manifestly Hermitian. 
        The bosons represent spin flip excitations above the classical ordered state.
        Next, a mean-field decoupling in real space is performed \cite{Takahashi_1989,Hida_1990} and the resulting bilinear Hamiltonian is solved by means of a Fourier and subsequent Bogoliubov transformation. 
        These transformations are applied to the total Hamiltonian including the quartic interaction terms yielding
            \begin{align}
                \label{eq:startHamil}
                \mathcal{H} = E^{(0)}_{\vphantom{1\leftrightarrow 1}} + 
                    \Gamma^{(0)}_{1\leftrightarrow 1} + \Gamma^{(0)}_{0\leftrightarrow 2} +
                    \mathcal{V}^{(0)}_{0\leftrightarrow 4} + \mathcal{V}^{(0)}_{1\leftrightarrow 3} + \mathcal{V}^{(0)}_{2\leftrightarrow 2}.
            \end{align}
						The subscripts $n\leftrightarrow m$ indicate the numbers of creation and annihilation operators of magnons 
						appearing in the labeled terms. Bilinear terms are denoted by $\Gamma$, quartic terms by $\mathcal{V}$. 
						For example, $\Gamma^{(0)}_{1\leftrightarrow 1}$ is given by
            \begin{align}
                \Gamma^{(0)}_{1\leftrightarrow 1} = \sum_{\vect{k}} \disp_0(\vect{k}) (\norord{\alp*{\vect{k}}\alp{\vect{k}}} + \norord{\bet*{\vect{k}}\bet{\vect{k}}})
            \end{align}
            where $\alpha^{(\dagger)}_{\vect{k}}$ and $\beta^{(\dagger)}_{\vect{k}}$ are the 
						bosonic magnon operators after the Bogoliubov transformation and
						$\omega_0(\vect{k})$ is the bare one-magnon dispersion.
            The colons $\norord{\ldots}$ indicate normal-ordered operators with respect to the mean-field ground state.
            The term $\Gamma^{(0)}_{0\leftrightarrow 2}$ comprises off-diagonal bilinear contributions that 
						are zero after the mean-field decoupling.
            The operators $\mathcal{V}^{(0)}_{n\leftrightarrow m}$ contain quartic terms 
						including the two-magnon interactions ($n=m=2$), but also couplings of the vacuum 
						to four-magnon states ($n=0,m=4$) and couplings of single-magnon states to three-magnon states ($n=1,m=3$).  
        The derivation and the final form of of all quartic contributions $\mathcal{V}^{(0)}$
				is given in App.\ \ref{a:aa-and-bb-terms}.
				
        Only certain kinds of quadratic and quartic operators respect the total spin conservation 
        \be
				S^{\text{tot}} = \sum_{i \in \Gamma_A} \abos*{i} \abos{i} - \sum_{i \in \Gamma_B} \bbos*{j} \bbos{j} 
				= \sum_k (\alp*{k} \alp{k} - \bet*{k} \bet{k}).
				\ee
        These terms are already present in the initial Hamiltonian at $\ell=0$ of the CST.
		Consequently, the CST-flow consists of the same set of operators for all three long-range ordered phases investigated in this work.
        On a technical level, the models differ only in their initial values for the prefactors of the operators and in their lattice symmetries which we will discuss in \Cref{ss:boundaryConditions}.
				
        After the CST flow induced by the $0n$-generator the effective Hamiltonian reads
        \begin{align}
            \label{eq:effHamil}
            \mathcal{H}_{\mathrm{eff}} = E + \Gamma_{1\leftrightarrow 1} + 
						\mathcal{V}_{1\leftrightarrow 3} + \mathcal{V}_{2\leftrightarrow 2}.
        \end{align}
        with the ground-state energy $E$ and the remaining quadratic terms $\Gamma_{1\leftrightarrow 1}$
				and quartic terms $\mathcal{V}_{1\leftrightarrow 3} + \mathcal{V}_{2\leftrightarrow 2}$.
        We stress again that the ground-state is decoupled from all 
				single- and multi-particle states by application of the $0n$-generator.
		
        In the code, we stop the flow when the residual off-diagonality (ROD), i.e., the square root of the sum over all squared entries of the generator $\eta$ \cite{Fischer_2010}, has dropped to values below $10^{-6}J$ in the bilayer model and below  $10^{-6}J_1$, respectively, in the \jj model.
        Note that the number of entries grows $\propto L^6$ resulting 
				in a increased numerical accuracy for higher $L$ because each residual term is reduced further.

        \subsection{Observables}
		\label{ss:observables}

        Here we discuss the observables and quantities used to study the quantum phase transition in all cases under consideration. These are the convergence of the CST flow itself, the sublattice magnetization, and the second derivative of the ground-state energy. The sublattice magnetization is the order parameter of the long-range ordered phases.

        If the basis change is induced by the $0n$-generator, a breakdown of the flow indicates a crossing of some excited energy with the ground-state energy and thus the collapse of the magnetically ordered phase.
        Hence, a first way to determine the critical point $\lambda_{\mathrm{c}}$ consists in locating the value of $\lambda$ where we first find a divergent flow. Here, $\lambda$ can be $\lambda_\perp$ or $\lambda_{12}$ depending on the model. For the bilayer model and the melting of the N\'eel order in the \jj model we consider increasing values of $\lambda$; for the melting of the columnar order we consider decreasing $\lambda$.
		In practice, we sample the parameter region close to the phase transition with a grid of width distance $\Delta\lambda = \num{0.001}$ and thereby determine $\lambda_{\rm c}$ with an uncertainty of $\Delta\lambda$.
         
        In addition, we determine two physical quantities from the ground-state energy per site that exhibit quantum critical power law behavior in the vicinity of a continuous phase transition.
        The first is the order parameter of the long-range ordered magnetic phases, i.e., the alternating magnetization.
        With the help of the Hellmann-Feynman theorem, we calculate the total sublattice magnetization by      
        \begin{align}
            M =  \frac{\mathrm{d}}{\mathrm{d}h_{\mathrm{alt}}} E_0(h_{\mathrm{alt} })   \bigg\vert_{h_{\mathrm{alt}}=0},
        \end{align}
        where we introduced an alternating magnetic field $h_{\mathrm{alt}}$ in the Hamiltonian favoring the long-range order by breaking the SU(2) symmetry explicitly.
        In practice, we compute the derivative as a ratio of differences $\left( E_0(h_{\mathrm{alt}})-E_0(0) \right)/ h_{\mathrm{alt}})$ for small $h_{\mathrm{alt}}= \num{1e-7} J$, cf.\ Refs.\ \cite{demid23,caci24}.
        To be able to compare different system sizes we consider the ground state energy per site $e_0$ and the sublattice magnetization per site $m$.
				
        Close to the quantum critical point $\lambda_{\rm c}$ of a continuous phase transition, 
		the sublattice magnetization $m$ displays a power-law behavior  
        \begin{align}
               m(\lambda) \propto \left(1-{\lambda/\lambda_c}\right)^\beta  
				\label{eqn:beta}
        \end{align}  
        with $\beta$ being the corresponding critical exponent.

        Moreover, we have access to the critical exponent $\alpha$ by means of  the second derivative of the ground-state energy per site with respect to  $\lambda$ which is a susceptibility towards changes of the control parameter. 
        Close to a continuous phase transition one has
            \begin{align}
                \frac{\mathrm{d}^2}{\mathrm{d}\lambda^2} e_0(\lambda) \propto \left(1-{\lambda/\lambda_c}\right)^{-\alpha} .
								\label{eqn:alpha}
            \end{align}  
        The second derivative is numerically evaluated by using the second order central difference scheme twice.
        In the columnar phase, an overall sign appears in the above brackets in Eqs. \eqref{eqn:beta} and \eqref{eqn:alpha} since the critical point is approached from large $\lambda$ towards smaller values of $\lambda$ leading to 
				$(1 - \lambda/\lambda_c)$ instead of $(\lambda/\lambda_c-1)$.

        \subsection{ Lattice discretization }
        \label{ss:boundaryConditions}

            In order to solve the CST flow numerically, the Brillouin zone (BZ) has to be discretized; we use the term ``Brillouin zone'' both for the magnetic Brillouin zone for the \jj model or the standard Brillouin zone for the Heisenberg bilayer, see below for details.
            To fulfill conservation of total momentum it is convenient to choose the mesh of sampling points equidistant.  All BZs considered in this work are rectangular in shape with $L \times L$ points where $L$ is the linear size.

            \begin{figure}
               \subfloat[\label{subfig:BZ_bilayer}]{%
                    \includegraphics[height=0.5\columnwidth]{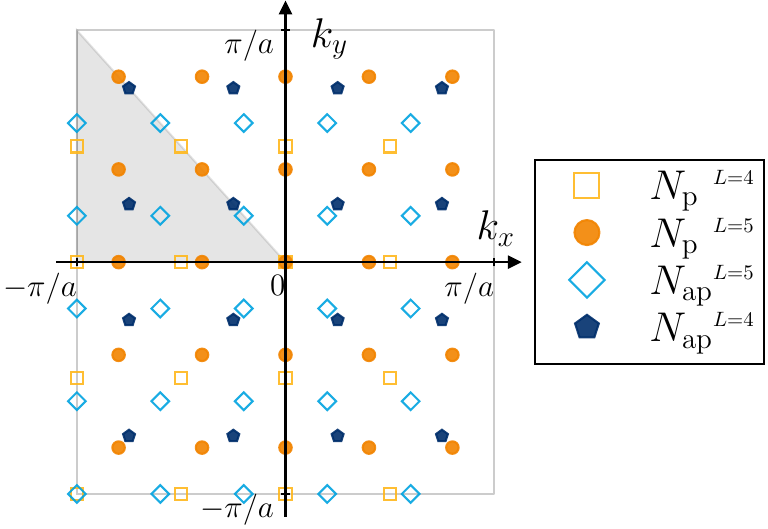}%
                }
                \\
                \subfloat[\label{subfig:MBZ_neel}]{%
                      \includegraphics[height=0.5\columnwidth]{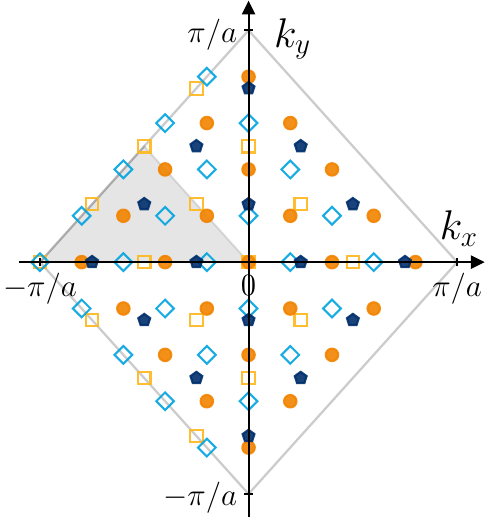}%
                }\hfill
                \subfloat[\label{subfig:MBZ_columnar}]{%
                      \includegraphics[height=0.5\columnwidth]{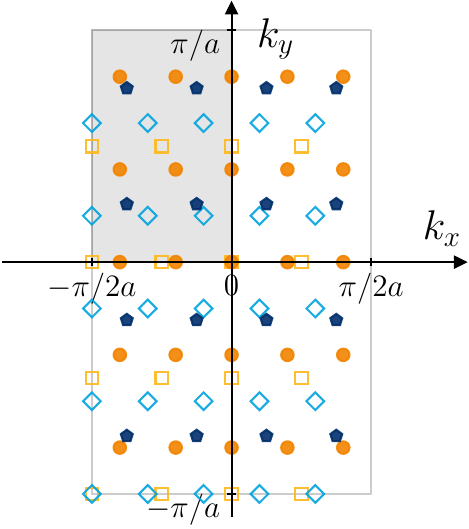}%
                }
                \caption{
                    Discretizations of the BZs for the three long-range ordered phases. 
										We show the two possible discretizations $\NAP$ and $\NP$ for a linear system size of $L=4(5)$.
                    The BZ of the Heisenberg bilayer is displayed in \Cref{subfig:BZ_bilayer}.
                    The  magnetic BZs of the two ordered phases of the \jj  model are shown in \Cref{subfig:MBZ_neel} for the Néel phase and  in \Cref{subfig:MBZ_columnar} for the columnar phase. 
                }
                \label{fig:boundaryConditions}
        \end{figure}

            Additionally, we adopt periodic or antiperiodic boundary conditions.
            For periodic boundary condition, a bosonic creation or annihilation operator is retrieved exactly after a shift by $L$ along one of the axes while they take a minus sign for the antiperiodic boundary conditions.
            This procedure helps to reduce finite size effects. 
            In momentum space, the difference between periodic and antiperiodic boundary condition manifests in sampling different points in the BZs, see \Cref{fig:boundaryConditions}.
            The antiperiodic conditions allows us to avoid the exact center $\Gamma = (0,0)$ where the mean-field solution displays an integrable divergence which make a numerical treatment on a discrete grid not feasible. Therefore, we set coefficients involving the $\Gamma$ point for the periodic boundary conditions to zero at the beginning of the flow. This approach was already taken in Refs.\ \onlinecite{Powalski15,Powalski18}.
            Since we remove a zero-dimensional point from a two-dimensional lattice, 
            we expect this error to vanish for large $L$, for instance in an appropriate extrapolation.
            Antiperiodic lattices automatically circumvent the problem by not sampling $\Gamma$.
	
            A certain disadvantage in $\Gamma$ not being part of the sampling points is that only even numbers of momenta can add to zero, i.e., comply with momentum conservation. For the models studied, displaying collinear order, this is not an issue because only even number of bosonic operators occur in each monomial.

            The use of different periodicity in the boundary conditions imply that the numerical results are not identical, even for the same $L$. But we expect that in the limit $L\to\infty$ the results coincide within numerical accuracy. We will use coincidence or the lack thereof as means to assess the accuracy of our results.
          In the following we will use the symbol $\NAP$ for results obtained with an antiperiodic boundary conditions and $\NP$ for the periodic counterpart.
          The present work is the first CST calculation showing $\NAP$ with even values of $L$ and $\NP$ with odd values of $L$. 						

            The BZs of the three different long-range ordered phases have different shape and point symmetries.
            In \Cref{fig:boundaryConditions}, we show the BZ of the Néel order in the Heisenberg bilayer (\Cref{subfig:BZ_bilayer}) as well as the magnetic BZs in the Néel (\Cref{subfig:MBZ_neel}) and the columnar phase (\Cref{subfig:MBZ_columnar}) of the \jj model.
            The different markers encode the various lattice discretizations $\NAP$ and $\NP$ and lengths $L=4,5$ used in the CST.
            Due to the point symmetry of the real space unit cell, only the shaded area needs to be taken into account for all integrals over the BZs and in the flow of the coupling constants.
            Note that both, the Néel phase in the Heisenberg bilayer model and in the \jj model shows an additional rotational symmetry, which can be used to reduce the number of couplings to be tracked, thereby enhancing the performance of the CST calculations.
						
            Although we always discretize with $L\times L$ sampling points, $L$ corresponds to different lengths in each phase because the unit cells of the three phases have different shapes.
            For the Heisenberg bilayer, both atoms are located in the center of the square unit cell with side length $a$, where $a$ is the lattice distance that we set to one for all models.
            Therefore, to know the length one needs to multiply $L$ by $a$. 
            The unit cell of the Néel phase of the \jj model is a diamond with length $\sqrt{2}a$ along the edges and therefore $L$ needs to be multiplies with $\sqrt{2}a$ to obtain lengths.
            Finally, the unit cell of the columnar phase is a $2a\times a$ rectangle.
            Therefore, $L$ has different length factors in $x$ and $y$ direction.  The smaller length scale is in units of $a$. These different lengths must be kept in mind when discussing the results in the following sections. 
    \section{Results}
    \label{s:results}
  
    In this section, we discuss the results obtained for the quantum melting of the three long-range ordered phases obtained from the described CST approach. 
    For all three cases we discuss the convergence of the CST flow, the ground-state magnetization, and the second derivative of the ground-state energy as described in \Cref{ss:observables}.
        The critical value and universality class of the unfrustrated Heisenberg bilayer on the square lattice are known and hence serve as a benchmark for the CST method.
        Then, we discuss our results for the breakdown of the Néel and the columner phase in the \jj  model and compare our results with the findings in literature.

        \subsection{Heisenberg bilayer}
        \label{ss:results:hbbilayer}
\begin{figure}[htb]
                \includegraphics[width=\columnwidth]{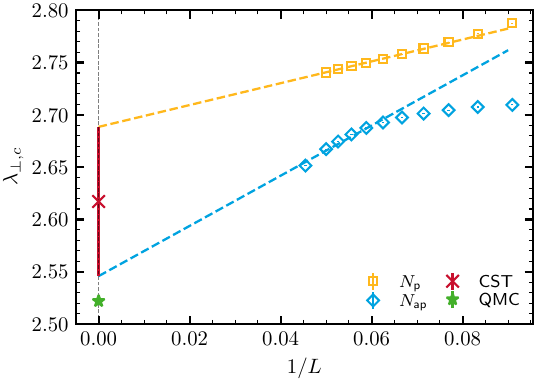}
                \caption{
                    Extrapolation of the last coupling for which the CST converges in the Heisenberg bilayer for different lengths $L$ and different boundary conditions.
                    The values of the last convergent $\lambda = J_{\perp}/J$ are shown.
                    The average of the linear fits for $\NP$ nd $\NAP$  is $\lambda^{\perp}_c = \num{2.62 \pm 0.07}$, which is $\approx 4\%$ above the literature value of $\num{2.5220 \pm 0.0001}$\cite{Wang_06,Lohoefer_15}.}
                \label{fig:results_hbb_lamc}
            \end{figure}

            In \Cref{fig:results_hbb_lamc}, we show $\lambda_{\perp,{\rm c}}$ as a function of $1/L$ for \mbox{$L \in (11, 20)$} and for all discretizations. 
            The values $\lambda_{\perp,{\rm c}}$ are the maximum values of $\lambda_{\perp}$ at which the CST flow is still converging. We see that the values for $\lambda_{\perp, \rm c}$ are larger than the value \mbox{$\lambda_{\perp,\rm c} = \num{2.5220 \pm 0.0001}$} in literature \cite{Wang_06,Lohoefer_15}, but are significantly closer to them than the results from linear spin-wave theory ($\lambda_{\perp, \rm c} \approx \num{13.6}$) and self-consistent spin-wave theory ($\lambda_{\perp,\rm c} \approx \num{4.2}$) \cite{Hida_1990}.

            The results for $N_{\text{p}}$ can be extrapolated to infinity with a linear fit in $1/L$.
            This yields $\lambda_{\perp, \rm c} \approx \num{2.7}$, which is about $7\%$ too large compared to the literature results.
            The results for $N_{\text{ap}}$ do not display a clear scaling behavior in $1/L$ for the accessible length scales.
            The critical points for small $L$ and $\NAP$ decrease slower than the results for $\NP$ which is expected due to the effects discussed in \Cref{ss:boundaryConditions}. 
            However, the values bend down strongly from $L \approx 15$ onwards.
            A linear fit of the last five values yields $\lambda_{\perp,\rm c} \approx \num{2.55}$ indicating that larger system sizes are needed for a reliable extrapolation.
            The average of the results of both boundary conditions is $\lambda_{\perp, \rm c} = \num{2.62 \pm 0.07}$.
            Interestingly, we do not observe the same behavior for the two ordered phases in the \jj  model, see below.
            Let us stress that the presented results are a significant improvement compared to any variant of spin-wave theory. At the same time, the breakdown of the flow is not able to quantitatively capture the critical values with the same accuracy as for example QMC.

\begin{figure}[htb]
                \includegraphics[height=0.65\columnwidth]{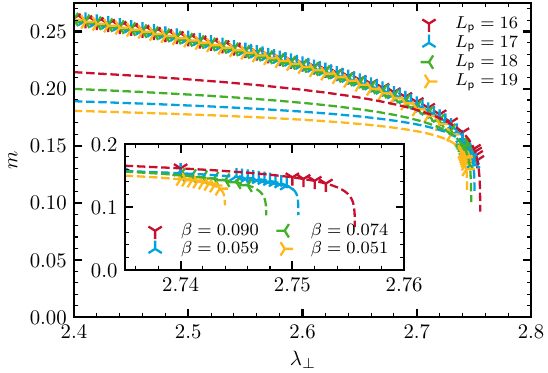}
                \caption{ 
                    The markers show the sublattice magnetization per site $m$ in the square lattice Heisenberg bilayer as a function of the control parameter $\lambda_{\perp}$ for different linear system sizes $L$. For clarity, the data shown is for $\NP$ only.
                    The dashed lines show the power-law fits using the last six values before the flow diverges.
                    This results in a trend towards too small values for the exponent $\beta$.}
                \label{fig:results_hbb_beta_raw}
            \end{figure}

            As discussed in \Cref{ss:observables}, we expect a power-law behavior of the sublattice magnetization $m$ and the second derivative of the ground-state energy per site $\mathrm{d}^2 e_0/\mathrm{d} \lambda^2$ in the vicinity of quantum phase transitions.
            Figure \ref{fig:results_hbb_beta_raw} shows such singular behavior for $m$ for values close to the critical point for $\NAP$ and for different $L$ together with power-law fits. 
            One sees clear signs of singularities in both quantities. 
            However, the last values for $m$ and $\mathrm{d}^2/\mathrm{d} \lambda^2 e_0$, where the CST flow is still converging, do not reach the expected values of $0$ or $-\infty$, respectively. 

            We also observe that the values for $\beta$ and $\alpha$ obtained from power-law fits on the last five points are by far too small and extrapolate to $0$ in the case of $\beta$ and to infinity for $\alpha$.
            Clearly, these findings are inconsistent with the known quantum-critical behavior of the square lattice Heisenberg bilayer. 
            This lead us to the conclusion that we do see the breakdown of the N\'eel order in the CST data, but we do not capture the critical behavior close to the phase transition in the divergence of the flow itself.
            One possible explanation is that neglected fluctuations due to the truncated hexatic or higher terms start playing a significant role close to the phase transition. 
            Additionally, a hallmark of continuous quantum phase transitions is the divergence of correlation lengths and hence fluctuations on larger length scales become increasingly more important.
            The finite discretization of the numerical solution of the CST flow might be limiting in this regard as well.

            Therefore, the power-law fits of the flow divergence are not the optimum quantity to extract quantum critical properties. 
            As a consequence, we choose a fit interval that is close to the phase transition, but is not yet so close that artifacts of the CST flow due to the truncation or due to the finite  lattice size  play a significant role.
            This approach is in the spirit of series expansions, where perturbative data obtained far from the critical point are extrapolated towards the critical point yielding often quantitative results. 
            As a criterion for this intermediate range of vicinity, we choose the principle of minimal sensitivity of the fit results on the fit interval \cite{steve81}. 
            This means, we chose the fit interval such that varying the upper and lower limit of the fit interval changes the result the least while staying as close to the phase transition as possible.
            The technical details of this approach are given in App.\ \ref{a:minimal_sensitivity}.

\begin{figure}[htb]
                \includegraphics[width=\columnwidth]{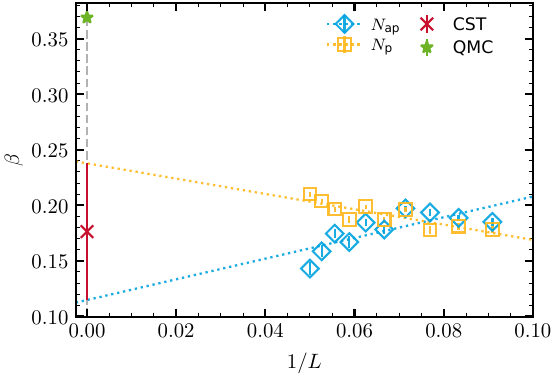}
                \caption{
                    Critical exponent $\beta$ determined from the sublattice magnetization in the square lattice Heisenberg bilayer for different linear system sizes $L$ and different boundary conditions.
                    The dotted lines show a linear fit for all lattice discretizations. 
                    The average of the linear fits is $\beta = \num{0.18 \pm 0.06}$ which is $\approx 52\%$ below the literature value of $\num{0.3689 \pm 0.0003}$.
                }
                \label{fig:results_hbb_beta}
            \end{figure}

            Figure \ref{fig:results_hbb_beta} shows the obtained values for $\beta$ found by the principle of minimal sensitivity for the fit interval and the boundary conditions $\NP$ and $\NAP$. Interestingly, the $\beta$ values for $\NAP$ and $\NP$ are very similar and $\NP$ does not show a strong dependence on $L$.
            The average of a linear extrapolations is $\beta\approx\num{0.18 \pm 0.06}$ which is significantly smaller than the literature value $\beta = \num{0.3689 \pm 0.0003}$ \cite{Campostrini_2002}. 
            However, we do find a finite value that is consistent within the different numbers of sampling points of the CST and significantly different from the mean-field value $\beta=\frac{1}{2}$ showing that we do capture non-trivial algebraic behavior. 
            Yet, our finding indicates that we do not have quantitative access to the critical behavior of $m$.

\begin{figure}
                \includegraphics[width=\columnwidth]{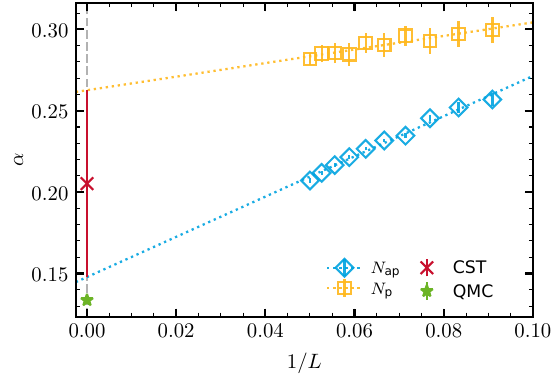}
                \caption{
                    Critical exponent $\alpha$ determined from the second derivative of the ground-state energy per site in the square lattice  Heisenberg bilayer for different system sizes $L$ and different boundary conditions.
                    The dotted lines show linear fits for all the data of one boundary condition. 
                    The average of the linear fits is $\alpha = \num{0.205 \pm 0.057}$ which lies $\approx 54\%$ above the literature value of $\num{0.1336 \pm 0.0015}$.
                }   \label{fig:results_hbb_alpha}
            \end{figure}

            Finally, \Cref{fig:results_hbb_alpha} shows the results obtained for the critical exponent $\alpha$ by the principle of minimal sensitivity of the fit interval for given $\NP$ and $\NAP$.
            The values for $\alpha$ lie on straight lines if plotted over $1/L$. The results for $\NP$ fall faster for  $1/L\to 0 $.
            By linear fits and averaging over $\NAP$ and $\NP$ we find $\alpha = \num{0.205 \pm 0.057}$ with a relative error of more than $\SI{25}{\percent}$ reflecting the large spread of the extrapolated results.
            The literature value is $\alpha=\num{ 0.1336 \pm 0.0015 }$ \cite{Campostrini_2002} which we strongly overestimate.
            We do not have convincing hypothesis for the relatively large difference between 
						the results of both boundary conditions. 
            The spread is similar to the spread in the values for the critical 
						$\lambda_{\perp, \rm c}$ in \Cref{fig:results_hbb_lamc}. 
            Note that for $\lambda_{\perp, \rm c}$  and $\alpha$, 
						the $\NAP$ results alone appear closer to the QMC results; for the exponent $\beta$, however,
						this is not the case. 
 
            In conclusion, the CST is able to determine the critical coupling with an accuracy of about $\SI{2}{\percent}$ by analyzing the divergence of the CST flow, but the calculations for $\NAP$ and $\NP$ differ significantly.
            Studying the critical exponents $\alpha$ and $\beta$ in the proximity of the divergence of the flow yields unphysical values which are most likely due to the finite size effects and truncation errors in the CST approach.
            An analysis based on the principle of minimal sensitivity \cite{steve81} of the fit interval allowed us to identify power-law behavior which is not of mean-field type.
            But the extracted values deviate strongly from the literature values.
	
        \subsection{\texorpdfstring{\jj}{TEXT} model}
        \label{ss:results:j1j2}

            As already discussed in \Cref{ss:boundaryConditions}, we are able to access both magnetically ordered phases of the \jj  model with the CST \and determine where these phases become unstable against quantum fluctuations. 
            This is done by studying the divergence of the flow in these phases.
            If the flow converges, it may still be that the phase is unstable against a first-order transition which does not result from diverging local fluctuations of the ground-state energy. 
            Hence, it may be that we overestimate the stability of the long-range ordered phases.

            We extrapolate $\lambda_{12,{\rm c}}$ as function of $1/L$ in both phases for $L\in(12, 20)$ and the two possible boundary conditions $\NP$, $\NAP$. As for the bilayer model, the values given are the average of the results of these two cases. In addition,  we again extract the critical exponents $\beta$ and $\alpha$ of the two phase transitions from the sublattice magnetization and the second derivative of the ground-state energy per site, as we did before in \Cref{ss:results:hbbilayer} for the bilayer model.

\begin{figure}[htb]
                \includegraphics[width=\columnwidth]{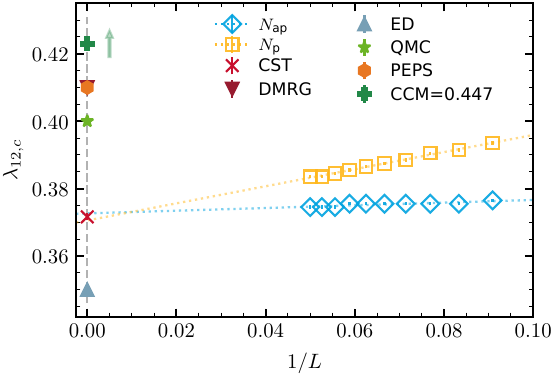}
                \caption{
                    Extrapolation of the last coupling for which the CST converges in the Néel phase of the \jj model for different lengths $L$ and different boundary conditions.
                    The average of the linear fits for $\NAP$ and $\NP$ is $\lambda_{12,{\rm c}} = \num{0.371 \pm 0.001}$, which is within $\approx 10\%$ in the range of the  literature values, except for CCM.
                }
                \label{fig:results_neel_lamc}
            \end{figure}

            Starting with the Néel phase in \Cref{fig:results_neel_lamc}, we find the critical value $\lambda_{12,{\rm c}}=\num{0.372 \pm 0.001}$.
            Here, the extrapolated values for $\NAP$ and $\NP$ agree very well, in contrast to what we had found in the Heisenberg bilayer.
            The values for $\NP$ fall faster for  $1/L\to 0$ which is likely due to the systematic omission of the wave vector $\Gamma=(0,0)$ for periodic boundary conditions as discussed in \Cref{ss:boundaryConditions}.
            Compared to the literature, the determined value of $\lambda_{12,{\rm c}}$ lies well within the results obtained by exact diagonalization (ED) with $\lambda_{12,{\rm c}}=\num{0.35}$ \cite{richterSpin1J1J22010}, variational QMC with $\lambda_{12,{\rm c}}=\num{0.4}$ \cite{moritaQuantumSpinLiquid2015}, and DMRG with $\lambda_{12,{\rm c}}=\num{0.41}$ \cite{jiangSpinLiquidGround2012}. 
            The deviations to these literature value are below $\SI{10}{\percent}$, in the range that the literature values scatter anyway.
            Only the results obtained by the CCM $\lambda_{12,c}=\num{0.447}$ deviates by about $\SI{17}{\percent}$.

\begin{figure}
                \includegraphics[width=\columnwidth]{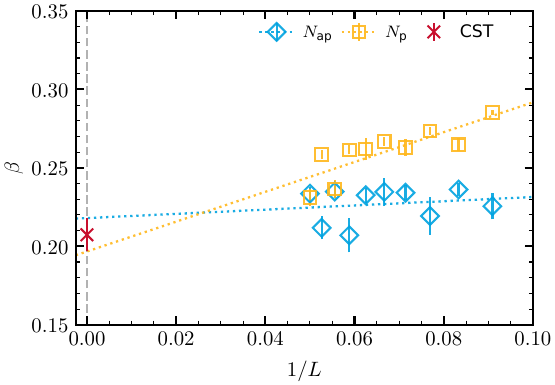}
                \caption{
                    Critical exponent $\beta$ determined from the sublattice magnetization in the Néel phase of the \jj  model for different system sizes $L$ and different boundary conditions.
                    The dotted lines show linear fits; the average of the extrapolated linear fits is $\beta = \num{0.21 \pm 0.01}$.
                }
                \label{fig:results_neel_beta}
            \end{figure}

            In \Cref{fig:results_neel_beta} the results for the exponent $\beta$ as a function of $1/L$ are shown which are again determined according to the principle of minimal sensitivity of the fit interval. 
            The values are a bit more scattered, but they indicate the range between $\num{0.2}$ to $\num{0.3}$.
            We find $\beta=\num{0.21\pm0.01}$ by linear extrapolation and averaging.
            We mention that in Ref.\ \onlinecite{moritaQuantumSpinLiquid2015} the authors deduced the much larger value $\beta\approx 0.5$. The system sizes, however, studied were not particularly large.

\begin{figure}
                \includegraphics[width=\columnwidth]{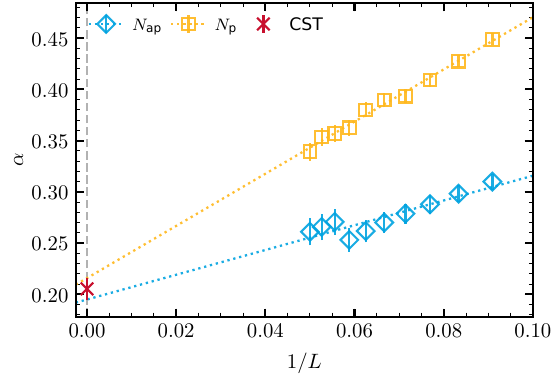}
                \caption{
                    Critical exponent $\alpha$ determined from the second derivative of the ground-state energy per site in the Néel phase of the \jj  model for different system sizes $L$ and different boundary conditions.
                    The dotted lines show linear fits; the average of the extrapolated linear fits $\alpha = \num{0.21 \pm 0.01}$.
                }
                \label{fig:results_neel_alpha}
            \end{figure}

            Continuing in the same manner as for $\beta$, \Cref{fig:results_neel_alpha} displays the results for $\alpha$ as function of $1/L$. 
            Similar to the cases before, we find a stronger  $L$ dependence of the data for $\NP$ which we attribute to the omission of the $\Gamma$ point. Yet, the linear extrapolations of the $\NP$ and the $\NAP$ data yield a very similar result.
            The average takes the value $\alpha=\num{0.21\pm0.01}$. It is interesting, that the $\alpha$ and the $\beta$ values appear to be identical. At present, this fact lacks an explanation.

\begin{figure}[htb]
                \includegraphics[width=\columnwidth]{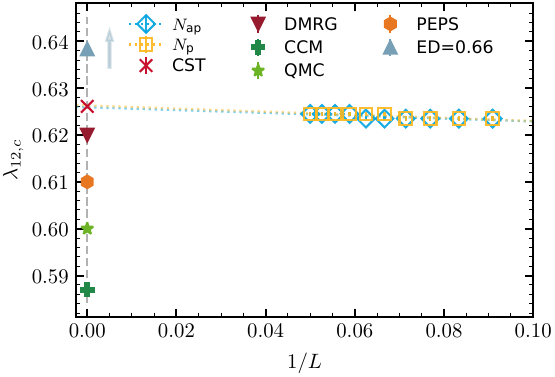}
                \caption{Extrapolation of the last coupling for which the CST converges in the columnar phase of the \jj model for different lengths $L$ and different boundary conditions.
                    The average of the linear fits for $\NAP$ and $\NP$ is $\lambda_{12, \rm c} = \num{0.6261 \pm 0.002}$ which deviates from the literature values by $\approx 6\%$ or less.
                }
                \label{fig:results_columnar_lamc}
            \end{figure}

            Finally, for the columnar phase we start again with the critical point of the phase transition $\lambda_{12,{\rm c}}$. In \Cref{fig:results_columnar_lamc}, the last converging couplings are depicted against $1/L$ for both boundary conditions. 
            Again, we observe that for smaller $L$ the stability of the magnetically ordered phase is overestimated resulting in a negative slope here for the linear extrapolation since the phase transition is approached from larger values of $\lambda_{12}$. 
            Remarkably, we see an excellent agreement of the $\NP$ and $\NAP$ data implying a small estimated error of extrapolation of the extrapolated and average value $\lambda_c=\num{0.6261\pm0.0002}$.
            Compared to ED results with $\num{0.66}$ \cite{richterSpin1J1J22010}, DMRG results with $\num{0.62}$ \cite{jiangSpinLiquidGround2012}, variational QMC results with $\num{0.6}$ \cite{moritaQuantumSpinLiquid2015}, and CCM results with $\num{0.587}$, the CST value fits very well.

\begin{figure}[htb]
                \includegraphics[width=\columnwidth]{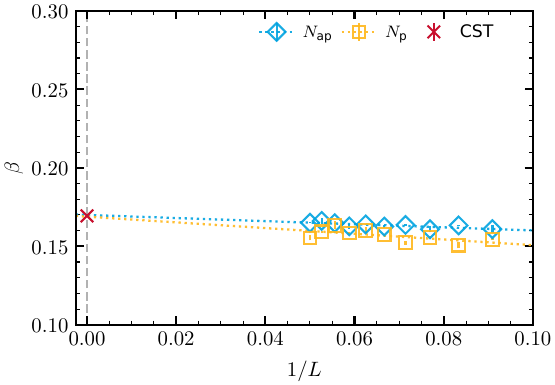}
                \caption{
                    Critical exponent $\beta$ determined from the sublattice magnetization in the columnar phase of the \jj  model for different system sizes $L$ and different boundary conditions.
                    The dotted lines show a linear fit for both boundary conditions. 
                    The average of the linear fits is $\beta = \num{0.169 \pm 0.001}$.
                } 
                \label{fig:results_columnar_beta}
            \end{figure}

\begin{figure}[htb]
                \includegraphics[width=\columnwidth]{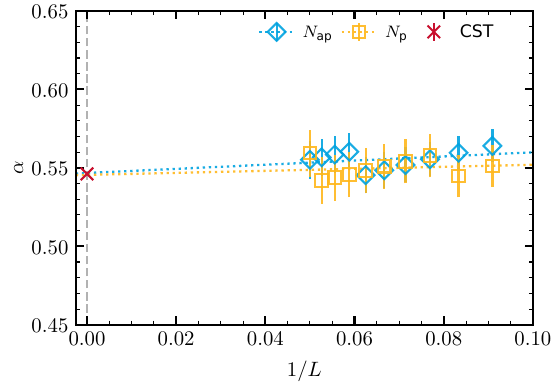}
                \caption{
                    Critical exponent $\alpha$ determined from the second derivative of the ground-state energy per site in the columnar phase of the \jj  model for different system sizes $L$ and different boundary conditions.
                    The dotted lines show a linear fit for both boundary conditions. 
                    The average of the linear fits is $\alpha = \num{0.5461 \pm 0.0006}$.
                    }
                    \label{fig:results_columnar_alpha}
                \end{figure}

            For the exponents $\beta$ and $\alpha$, the results for $\NAP$ and $\NP$ also agree very nicely.
            These values have been determined again by the method of minimal sensitivity described earlier, see also App.\ \ref{a:minimal_sensitivity}.
            The obtained values are $\beta = \num{0.169 \pm 0.001}$, see \Cref{fig:results_columnar_beta}, and $\alpha = \num{0.5461 \pm 0.0006}$, see \Cref{fig:results_columnar_alpha}.

            As discussed in \Cref{ss:j1j2model}, the nature of the phase transitions in the \jj  model is still under debate. No comparison of the exponents with the literature values is possible.
            The combination of $\beta$ and $\alpha$ for both phase transitions in the \jj  model do not fit to a universality class that we are aware of although we could extrapolate them quite robustly.
            The results for the critical values are close to the values in  literature.

\begin{figure}[htb]
                \includegraphics[width=\columnwidth]{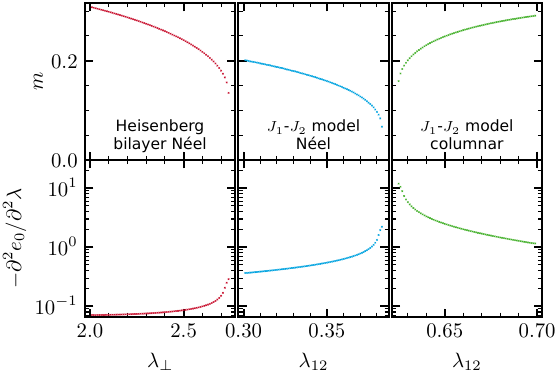}
                \caption{
                    The upper panels show the sublattice magnetization per site $m$ for the three different 
										long-range ordered phases. The lower panels show the corresponding second derivative of the 
										ground-state energy per site $\partial^2 e_0 / \partial \lambda^2$ on a logarithmic scale.
                    In all three cases, results for the antiperiodic boundary conditions $\NAP$
										are shown for linear system size $L=20$.
                    }   
                \label{fig:results_comparision}
            \end{figure}

            In contrast to the square lattice Heisenberg bilayer, we observe a remarkably good agreement of 
						the results from the two boundary conditions $\NP$ and $\NAP$. 
            Additionally, we want to point out that the CST captures the singularity of $m(\lambda_{12})$ and 
						of $\partial^2 e_0 / \partial \lambda^2$ to a much larger extent than for the Heisenberg bilayer.
            The magnetization can be traced to lower values closer to zero and the second derivative of $e_0$ 
						can also be traced to much lower values, as shown for example in \Cref{fig:results_comparision} 
						for $\NAP$ and $L=20$. This fact can be a hint, that the truncation scheme based on the scaling 
						dimension is better suited for the phases in the \jj  model than for the square lattice Heisenberg bilayer.

            Another interesting observation lies in the fact that the critical exponents of the two phase transitions in the \jj model are different, i.e., the $\alpha$ exponent of the melting of the N\'eel phase differs from the $\alpha$ exponent of the melting of the columnar phase; similarly for the $\beta$ exponent. This can be seen as evidence for at least
						two intermediate phases, one adjacent to the N\'eel order and one adjacent to the columnar order.
						But it appears to us that this argument is not compelling since the critical exponents are also depend
						on the symmetries of the melting phases and these are not identical for the N\'eel and the columnar ordered
						phase.
            
    \section{Conclusion}
    \label{s:conclusion}

        We applied the continuous similarity transformations (CSTs) in momentum space to long-range ordered magnetic phases in the unfrustrated square lattice Heisenberg bilayer and in the strongly frustrated \mbox{\jj model}. The CST was based on the Dyson-Maleev representation of the spins with a truncation in scaling dimension $\dsc$ keeping all operators with $\dsc\le2$.
        While the CSTs  have been applied successfully to the N\'eel ordered phase of the square lattice Heisenberg model as well as to the symmetry-enhancement transition in the square lattice XXZ model, this work is the first application to quantum phase transitions out of long-range ordered phases.

        The quantum melting of the long-range ordered phases is studied by the convergence of the CST flow for the $0n$ generator which aims at separating the ground state from all other excited states.
        In addition, the power-law behavior of the alternating magnetization per site with critical exponent $\beta$ and of the second derivative of the ground-state energy per site with critical exponent $\alpha$ is analyzed.
        Technically, we extended the types of discretizations  of the relevant Brillouin zones exploiting perioidic and antiperiodic boundary conditions with even and odd linear system size $L$.

        For the breakdown of the N\'eel phase of the square lattice Heisenberg bilayer we obtain for $\lambda_{\perp}=J_\perp/J$ the critical value $\lambda_{\perp, \rm c} = \num{2.66 \pm 0.03}$ which lies $\approx 5\%$ above the literature value of $\num{2.5220 \pm 0.0001}$ \cite{Wang_06}.
        This marks a significant improvement over self-consistent mean-field theory with $\lambda_{\perp, \rm c} = \num{4.2}$, but the extrapolations of the results from periodic and antiperiodic boundary conditions do not agree well and hinder a more reliable   determination of the critical coupling.

        Our analysis of the sublattice magnetization and the second derivative of the ground-state energy show clear evidence of singular behavior and hence of the onset of a quantum phase transition. But the CST for the models expressed in terms of magnons breaks down before the sublattice magnetization vanishes. 
        This leads us to the conclusion that the values in direct vicinity of the critical point are influenced by artifacts of the divergence itself and do not display the genuine critical behavior.
        Such artifacts can be due to the finite size of the discretized Brillouin zone and/or the truncation of terms beyond scaling dimension $\dsc=2$.
        
        We presented an algorithm based on the principle of minimal sensitivity to find ranges of the couplings that are close to the phase transitions, but not yet dominated by these artifacts, yielding consistent values for the critical exponents for different boundary conditions and lattice sizes.
        The resulting values are $\beta = \num{0.240 \pm 0.003}$ (lit $\num{0.3689 \pm 0.0003}$ \cite{Campostrini_2002}) and $\alpha = \num{0.205 \pm 0.057}$ (lit $\num{ 0.1336 \pm 0.0015 }$ \cite{Campostrini_2002}).
        This illustrates that critical exponents could not yet be reproduced by the CSTs in their current set up. Yet the CST values constitute a clear improvement over mean-field behavior.

        Due to the strongly frustrated nature of the \jj  model, critical points as well as the nature of the phase transitions and their universality classes associated with the quantum melting of the Néel and the columnar phase are not known precisely to date.
        Assuming that the phase transitions are continuous, the CST analysis yields $\lambda_{12,{\rm c}}=\num{0.371 \pm 0.001}$, $\beta=\num{0.20\pm0.02}$, and $\alpha=\num{0.21\pm0.01}$ for the melting of the Néel phase and $\lambda_c=\num{0.6261\pm0.0002}$, $\beta = \num{0.169 \pm 0.001}$, and $\alpha = \num{0.5461 \pm 0.0006}$ for the melting of the columnar phase.
        We point out that the CST results may overestimate the stability of the long-range ordered phases if the true phase transitions are first order.

        Yet, the critical CST values agree well with the ones from other numerical methods.
        Interestingly, for the \jj  model, the extrapolated values for $\lambda_{12,{\rm c}}$, $\beta$, and $\alpha$ from the two boundary conditions agree very nicely, while this agreement is rather poor for the square lattice Heisenberg bilayer. This comes as a surprise since the bilayer system is well established for displaying a continuous phase transition.
	We attribute this differing behavior to the smaller length scales treated in 
	the Heisenberg bilayer as well as an indication that the representation by Dyson-Maleev magnons 
	is better suited for the \jj model than for the square lattice Heisenberg bilayer.

        Finally, we conclude that the CST significantly improves the results of self-consistent mean-field theory and is able to locate quantum critical points up to a few percent. The precision for the critical exponents is difficult to assess since extrapolations do not agree so well for the system where the exponents are known while the consistent extrapolations do not yield exponents of an established universality class.

        Concerning the used truncation of the CST flow, an improved treatment would require to go beyond quartic operators, tracking hexatic terms with five independent momenta instead of three.
        This appears to be computationally not feasible even for moderate system sizes to date.
        A realistic option could be to select a finite number of hexatic terms by a heuristic argument.

        An interesting  direction of future research is to change the spin representation.
        The Dyson-Maleev transformation is designed to describe small fluctuations around a state with long-range magnetic order. Thus, it might not be optimally suited to describe the melting of this order precisely at the quantum critical points.
        A Schwinger boson representation \cite{auerb88} can capture both, ordered and disordered quantum phases, and could therefore be a better starting point.

        \begin{acknowledgments} 
            We gratefully acknowledge financial support by the Deutsche Forschungsgemeinschaft (DFG, German Research Foundation) through projects UH 90-14/1 (DBH/GSU), and SCHM 2511/13-1 (MRW/KPS). 
            We thankfully acknowledge HPC resources provided by the Erlangen National High Performance Computing Center (NHR@FAU) of the Friedrich-Alexander-Universit\"at Erlangen-N\"urnberg (FAU).
            KPS acknowledges further financial support by the German Science Foundation (DFG) through the Munich Quantum Valley, which is supported by the Bavarian state government with funds from the Hightech Agenda Bayern Plus.
        \end{acknowledgments}

%

    \FloatBarrier 

    \begin{appendix}
		
        \section{Self-consistent mean-field theory for intra-sublattice interactions}
        \label{a:aa-and-bb-terms}

        In this appendix we present the derivation of the self-consistent mean-field theory based on the Dyson-Maleev transformation for the Heisenberg interaction between pairs of spins on the same sublattice A or B. 
        The solution of the self-consistent mean-field theory provides the initial values for the CST in momentum space.

        We presented a detailed description of the analogous steps for inter-sublattice interactions in the XXZ model on the square lattice in the appendix of Ref.~\onlinecite{Walther23}.
        In that appendix, the interaction between a spin on sublattice $A$ with a spin on sublattice $B$ was derived.
        The results for the inter-sublattice interaction needed in the present paper is identical to the results in Ref.~\onlinecite{Walther23} except for two minor adaptions: 
        First, the sum over distance vector $\delta$ between nearest neighbors in the XXZ model has to be adapted to the relevant NN distances in the model at hand.
        Second, we do not study spin anisotropy in the present paper.
        Therefore, the anisotropy parameter can be set to one in the models discussed here.
        Since the inter-sublattice terms are already known, we focus in this appendix on the discussion of intra-sublattice terms.

        First, we can restrict ourselves to the explicit discussion of the intra-sublattice terms $S^{(A)}_{i}S^{(A)}_{j}$ because the $S^{(B)}_{i}S^{(B)}_{j}$  terms can then be deduced  easily since the Hamiltonian and the assumed order are invariant under spin flip and the translation by a lattice vector. 
        Due to this symmetry, the final form of the $S^{(B)}_{i}S^{(B)}_{j}$ terms can be inferred by replacing $\abos{i} \leftrightarrow \bbos{i}$ and a subsequent Hermitian conjugation.
        We use this symmetry as well in the numerical solution of the flow equations to minimize the number of  terms that we have to track. 

        Second, we base the derivation on the generic term
        \bes
        \begin{align}
                \mathcal{H}^{AA} &= \sum_{i}\sum_{\delta \in \vect{\delta}} 
								\vect{S}^{(A)}_i \vect{S}^{(A)}_{i + \delta} \\
								&= \sum_{i}\sum_{\delta \in \vect{\delta}}  \Big[S_{i}^z S_{i+ \delta}^z +  
								\frac{1}{2}\left(  S_{i}^+ S_{i+ \delta}^- +  S_{i}^- S_{i+ \delta}^+ \right)\Big], 
								\label{eq:app_hb}
        \end{align}
        \ees
        where $i$ labels all unit cells and \vect{\delta} labels the distances from one unit cell to the other ones.
        The exact form of \vect{\delta} is irrelevant for the computation and we can leave it as a variable here. 
        Eventually, the actual distances for the different interactions present in the system will be inserted.
        Potential double counting of interactions must be compensated by appropriate prefactors. 
        In \Cref{eq:app_hb} we introduced the spin raising (lowering) operator $S^+$ ($S^-$) and dropped the superscript $(A)$ since all operators are taken  to be located on sublattice $A$ in this calculation.
       
        We use the Dyson-Maleev representation where the elementary excitations of the system,  i.e., spin flips for $S=1/2$, relative to the classical Néel state are represented by the creation of a bosons.
        The Néel state serves as the vacuum for these bosons.
        For the $A$ sublattice, we use the from
        \bes
            \begin{align}
                S^z_i &= S - \abos{i} \abos*{i},\\
                S_i^- &= \sqrt{2S} \abos*{i},\\
                S_i^+ &= \sqrt{2S}\left[ 1 - {\abos*{i} \abos{i}/(2S)}\right]\abos{i},
            \end{align}
        \label{eq:ap:dysonMaleevA}
        \ees						
        where $a_i^{(\dagger)}$ are bosonic annihilation (creation) operators.

        For the $B$ sublattice, we use the Dyson-Maleev representation in the variant
		\bes
        \begin{align}
                S^z_i &= -S + \bbos{i} \bbos*{i},\\
                S_i^- &= \sqrt{2S} \bbos{i},\\
                S_i^+ &= \sqrt{2S}\bbos*{i}\left[ 1 - {\bbos*{i} \bbos{i}/(2S)}\right],
        \end{align}
        \label{eq:ap:dysonMaleevB}
        \ees
        where again $b_i^{(\dagger)}$ are bosonic operators acting on the $B$ sublattice.
        Choosing the Dyson-Maleev representations in this way limits the resulting operators of the inter-sublattice interaction to bilinear and quartic terms.

        Inserting \Cref{eq:ap:dysonMaleevA} into Eq.~\eqref{eq:app_hb} yields
        \begin{align}
                \mathcal{H}^{AA} &= -J \sum_{i, \delta \in \vect{\delta}} \Big[ S^2 \nonumber\\
                &\quad  + S \left( - \abos*{i} \abos{i} - \abos*{i+\delta} \abos{i+\delta} 
								+ \abos*{i}\abos{i+\delta} + \abos{i}\abos*{i+\delta} \right) 
								\nonumber\\
                &\quad + \abos*{i}\abos{i}\abos*{i+\delta}\abos{i+\delta} 
								\nonumber\\
                &\quad - \frac{1}{2} \left( \abos*{i}\abos{i}\abos{i} \abos*{i+\delta}  
                + \abos*{i}\abos*{i+\delta}\abos{i+\delta} \abos{i+\delta} \right)
                \Big].
								\label{methods:eq:dm-aa-interations}
	      \end{align}
        Next, we perform a standard mean-field decoupling \cite{Takahashi_1989}
        \be
                  a^{(\dagger)}_i \tilde{a}^{(\dagger)}_j =  
                                  \norord{ a^{(\dagger)}_i \tilde{a}^{(\dagger)}_j} +  
                                  \cexpval{ a^{(\dagger)}_i \tilde{a}^{(\dagger)}_j },
        \ee
        where $\norord{...}$ indicates normal-ordered operators and $\cexpval{ ... }$ are the vacuum expectation values. 
		For quartic terms, we use Wick's theorem to express all terms by their normal-ordered expressions.
				
        The conservation of the total spin component
		\be
            S^z_{\rm tot} = \sum_{i} (S^z_{(A),i} + S^z_{(B),i})
            = \sum_{i} (\abos*{i}\abos{i} - \bbos*{i}\bbos{i})
				\ee
        allows us to infer
        \be
            \cexpval{ \abos{i} \abos{i} } = \cexpval{ \abos*{i} \abos*{i} } = 0,
        \ee
        which simplifies the results of the normal ordering. Additionally, we define
        \bes
        \begin{align}
            n &\coloneqq \cexpval{ \abos*{i} \abos{i} } = \cexpval{ \abos*{i+\delta} \abos{i+\delta} } \\
            t &\coloneqq \cexpval{ \abos*{i+\delta} \abos{i} } = \cexpval{ \abos*{i} \abos{i+\delta} }
            \label{eq:app_nandt}
        \end{align}
        \ees

        Collecting all terms we finally obtain
        \begin{equation}
            \begin{aligned}
                \mathcal{H}^{AA} = & -J \sum_{i, \delta \in \vec{\delta}} 
								\Big[   E_{0,AA} + \tilde{A} \norord{ \abos*{i} \abos {i} } 
								+ \tilde{B} \norord{ \abos*{i} \abos {\idel} } \\
                & + \norord{ \abos*{i}\abos{i}\abos*{\idel}\abos{\idel} } \\
                & - \frac{1}{2} \left( \norord{ \abos*{i}\abos{i}\abos{i} \abos*{\idel} } 
								+ \norord{ \abos*{i}\abos*{\idel}\abos{\idel} \abos{\idel} } \right) \Big]
            \end{aligned}
            \label{eq:app_normalorders}
        \end{equation}
        with
        \bes
        \begin{align}
            E_{0,AA} &= 
                 -S^2 + 2S \left(  -n + t\right) 
                + n^2 + t^2 - 2 n t \\
            \tilde{A} &=2 \big( -S + n - t \big)\\
            \tilde{B} &=2 \big( S + t -  n \big)\,.
        \end{align}
        \ees

        Next, we exploit the discrete translation symmetry by performing a Fourier transformation to a reciprocal space.
        The only normal ordered operators conserving $S^z_{\rm tot}$ are $\norord{ \abos*{k} \abos {k} }$ and $\norord{ \abos*{1}\abos*{2}\abos{3}\abos{4}}$.
		Eventually, we obtain
        \begin{align}
                \mathcal{H}^{AA} &= -J \sum_{i, \delta \in \vec{\delta}} E_{0,AA} -J \sum_{k} 
								\norord{ \abos*{k} \abos {k} } \tilde{A}_k \nonumber\\
                & -\frac{J}{N} \sum_{1,2,3,4} \delta(1+2-3-4) \norord{ \abos*{1}\abos*{2}\abos{3}\abos{4}} 
								\widetilde{\gamma}(2,3,4),
            \end{align}
        where we used
        \bes
        \begin{align}
				     \gamma(k) &\coloneqq \sum_{\delta \in \vect{\delta}} \eee^{- \ii k \delta}\\
            \tilde{A}_k &\coloneqq \left( \tilde{A} + \gamma(k) \tilde{B} \right) \\
						\widetilde{\gamma}(2,3,4) &\coloneqq \gamma(2-4) - \frac{\gamma(2)}{2} - \frac{\gamma(2-3-4)}{2}.
        \end{align}
        \ees

        The next step is to perform a Bogoliubov transformation considering the full Hamiltonian with all contributions, i.e.,  not only the intra-sublattice terms.
        The full Hamiltonian in momentum space takes the form
        \begin{equation}
            \begin{aligned}
                \mathcal{H} &= E_0  \\ 
                &+ \sum_{\vect{k}\in \text{MBZ}} A_{\vect{k}} \big( \norord{ \abos*{\vect{k}} 
                   \abos{\vect{k}} } + \norord{\bbos*{\vect{k}} \bbos{\vect{k}}} \big) 
                                                      \\
                &+ \sum_{\vect{k}\in\text{MBZ}} B_{\vect{k}} 
								 \big( \norord{ \abos{\vect{k}} \bbos{-\vect{k}}}
                 + \norord{\abos*{\vect{k}}\bbos*{-\vect{k}}} \big)  
                                                        \\
                &+ \widetilde{\mathcal{V}},
            \end{aligned}
				    \label{eq:app_hamilton}
        \end{equation}
        with coefficients $A_k$ and $B_k$ for the bilinear terms and the quartic terms $\widetilde{\mathcal{V}}$.
        The $AA$ terms contribute to $E_0$ via $E_{0,AA}$, to $A_k$ via $\tilde{A}_k$,  
		and to $\widetilde{\mathcal{V}}$ via the terms $\abos*{1}\abos*{2}\abos{3}\abos{4}$.
        The analogous relations hold for the $BB$ terms.

        The bilinear part of $\mathcal{H}$ is diagonalized by the Bogoliubov transformation of the form
        \bes
        \begin{align}
            \abos*{\vect{k}} &= l_{\vect{k}} \alp*{\vect{k}} + m_{\vect{k}} \bet{-\vect{k}} \\
                    \bbos*{\vect{k}} &= m_{\vect{k}} \alp{\vect{-k}} + l_{\vect{k}} \bet*{\vect{k}} \\
             \text{with} \quad  1&= l_{\vect{k}}^2 -  m_{\vect{k}} ^2 = 1.
			  \end{align}
        \ees
        We parametrize explicitly as in Refs.\ \cite{Powalski15, Powalski18, Uhrig_2013}
        \bes
            \begin{align}
                 \mu_{\vect{k}} &\coloneqq \sqrt{1 - \left({B_{\vect{k}}/A_{\vect{k}}} \right)^2 } \\
                 l_{\vect{k}} &\coloneqq \sqrt{ \frac{1- \mu_{\vect{k}}}{2  \mu_{\vect{k}}} }
									\\
                 m_{\vect{k}} &\coloneqq - \sign(\gamma(\vect{k})) 
                                 \sqrt{ \frac{1 + \mu_{\vect{k}}}{2  \mu_{\vect{k}}} } 
                                  \\
                 x_{\vect{k}} &\coloneqq \sign(\gamma(\vect{k})) 
								\sqrt{ \frac{1+ \mu_{\vect{k}}}{1- \mu_{\vect{k} } } }.             
             \end{align}
        \ees
		Note that $m_{\vect{k}}= - x_{\vect{k}} l_{\vect{k}}$ holds.
				
        Following these steps, we rewrite the  Hamiltonian \eqref{eq:startHamil} in the form
        \begin{align}
            \mathcal{H} = E^{(0)} + \Gamma^{(0)} + \mathcal{V}^{(0)}
        \end{align}
        with 
        \begin{align}
            \Gamma^{(0)} &= \Gamma^{(0)}_{1\leftrightarrow 1} + \Gamma^{(0)}_{0\leftrightarrow 2} \\
            \mathcal{V}^{(0)} &= \mathcal{V}^{(0)}_{0\leftrightarrow 4} + \mathcal{V}^{(0)}_{1\leftrightarrow 3} + \mathcal{V}^{(0)}_{2\leftrightarrow 2}
        \end{align}
        where the term $\Gamma^{(0)}$  the quadratic terms and $\mathcal{V}^{(0)}$  all the quartic terms.
        $\mathcal{V}^{(0)}$ can be written as
        \begin{align}
            \mathcal{V}^{(0)} &= \frac{JZ}{N} \sum_{1234} \delta(1+2-3-4) l_{1} l_{2} l_{3} l_{4} \Big( \nonumber\\
            &
            V_{1234}^{(\rm a)} \alp*{1} \alp*{2} \alp{3}   \alp{4} +
            V_{1234}^{(\rm b)} \alp*{1} \alp*{2} \alp{3}   \bet*{-4} + \nonumber\\
            &
            V_{1234}^{(\rm c)} \alp*{1} \alp{2}  \bet*{-3} \bet*{-4} +
            V_{1234}^{(\rm d)} \alp*{1} \alp{-2} \alp{3}   \bet{4} + \nonumber\\
            &
            V_{1234}^{(\rm e)} \alp*{1} \alp{-2} \bet*{-3} \bet{4} +
            V_{1234}^{(\rm f)} \alp*{1} \bet*{2} \bet*{-3} \bet{4} +\nonumber\\
            &
            V_{1234}^{(\rm g)} \alp{-1} \alp{-2} \bet{3}  \bet{4} +
            V_{1234}^{(\rm h)} \alp{-1} \bet*{2} \bet{3}  \bet{4} + \nonumber\\
            &
            V_{1234}^{(\rm i)} \bet*{1} \bet*{2} \bet{3} \bet{4} 
            \Big).
        \end{align}
			
        The $AA$ interactions contribute to each vertex function $V^{(i)}_{1234}$
        according to
        \bes
        \begin{align}
            V_{(AA),1234}^{(\rm a)} &= \frac{1}{4}
                \ v_1(1,2,3,4)
            \\
            V_{(AA),1234}^{(\rm b)} &= -\frac{1}{2}x_{4}\
                \ v_1(1,2,3,4)
            \\
            V_{(AA),1234}^{(\rm c)} &= \frac{1}{4}x_{3}x_{4}
                \ v_1(1,2,3,4)
            \\
            V_{(AA),1234}^{(\rm d)} &= -\frac{1}{2}x_{4}
                \ v_2(1,2,3,4)
            \\
            V_{(AA),1234}^{(\rm e)} &= x_{3} x_{4}
                \ v_2(1,2,3,4)
            \\
            V_{(AA),1234}^{(\rm f)} &= -\frac{1}{2}x_{2}x_{3}x_{4}
                \ v_2(1,2,3,4)
            \\
            V_{(AA),1234}^{(\rm g)} &= \frac{1}{4}x_{3}x_{4}
                \ v_3(1,2,3,4)
            \\
            V_{(AA),1234}^{(\rm h)} &= -\frac{1}{2}x_{2}x_{3}x_{4}
                \ v_3(1,2,3,4)
            \\
            V_{(AA),1234}^{(\rm i)} &= \frac{1}{4}x_{1}x_{2}x_{3}x_{4}
                \ v_3(1,2,3,4)
        \end{align}
        \ees
        where we defined 
        \bes
        \begin{align}
            v_1(1,2,3,4) &\coloneq \widetilde{\gamma}(1,3,4)+\widetilde{\gamma}(1,4,3)+ \nonumber\\
            & \widetilde{\gamma}(2,3,4)+\widetilde{\gamma}(2,4,3)\\
            v_2(1,2,3,4) &\coloneq \widetilde{\gamma}(1,-2,3)+\widetilde{\gamma}(1,3,-2)+\nonumber\\
            &\widetilde{\gamma}(-4,-2,3)+\widetilde{\gamma}(-4,3,-2) \\
            v_3(1,2,3,4) &\coloneq \widetilde{\gamma}(-3,-1,-2)+\widetilde{\gamma}(-3,-2,-1)+\nonumber \\
            &\widetilde{\gamma}(-4,-1,-2)+\widetilde{\gamma}(-4,-2,-1)
        \end{align}
        \ees 
       Note that we symmetrized the contributions in order to account for the symmetry of swapping two wave vectors that are the arguments of the same creation or annihilation operators. 
        For example, $\alp*{1} \alp*{2} \alp{3}   \alp{4}$ is identical to $\alp*{2} \alp*{1} \alp{3} \alp{4}$ and this should also be reflected in the coefficients $V^{(a)}_{(AA),1234}$. 
		The analogous contributions of $S^{(B)}_iS^{(B)}_j$ can again be inferred by swapping $\alpha \leftrightarrow \beta$ in combination with a Hermitian conjugation.

        Finally, $n$ and $t$ are determined by performing the Fourier and Bogoliubov transformation on the right hand side of Eq.~\eqref{eq:app_nandt} and evaluating the vacuum expectation values. 
		This yields the self-consistency equations
        \bes
        \begin{align}
            n &= \frac{1}{N} \sum_{\vect{k}} \cexpval{ \abos*{\vect{k}} \abos{\vect{k}} } 
                                =\frac{1}{N} \sum_{\vect{k}} l_{\vect{k}}^2 \\
            t &= \frac{1}{N} \sum_{\vect{k}}\gamma(\vect{k}) 
                            \cexpval{ \abos*{\vect{k}} \abos{\vect{k}} } 
                        = \frac{1}{N} \sum_{\vect{k}} \gamma(\vect{k}) l_{\vect{k}}^2.
        \end{align}
        \ees
        They are solved numerically by iteration to convergence by a Gauss-Kronrod quadrature of order 601 as implemented in the ``boost C++ libraries'' project.
		Since $l_k$ is divergent for $k\rightarrow(0,0)$, we exclude wave vectors with $\abs{k}< 10^{-8}$ in the integration.
        The iteration is stopped when the change of each parameter is smaller than $10^{-13}$.
        If the $AA$ and $BB$ interactions are different, it may happen that the $t$ values differ because different  $\vect{\delta}$ and hence  different $\gamma(\vect{k})$ need to be considered. Then, these $t$ values have to be solved separately.

  \section{Power-law fits for critical behavior and exponents}
    \label{a:minimal_sensitivity}
		
        In this appendix, we discuss in detail how we extracted the critical exponents for the long-range ordered phases investigated in the main text. 
        Let us consider an observable which displays singular behavior near the breakdown of the CST flow, e.g., the second derivative of the ground-state energy or the sublattice magnetization.    
        Taking only the values in the immediate proximity of the breakdown into account, we found non-physical behavior of the critical exponents in all considered cases.
        To circumvent these artifacts, we set up a method to choose the fit interval of the coupling as close as possible to the critical point, but not too close either to avoid spurious effects of the divergence of the CST flow.

        The values of the observable are computed on an equidistant grid of the coupling.
        For the Néel phase of the square lattice Heisenberg bilayer we used a grid with spacing $\Delta_{\lambda_\perp}=0.01$ and for both, the Néel and the columnar phase of the \jj  model, a grid with spacing $\Delta_{\lambda} = 0.001$. 

        \begin{figure}[htb]
            \centering
            \subfloat[\label{subfig:beta_fit}]{%
            \includegraphics[width=0.99\columnwidth]{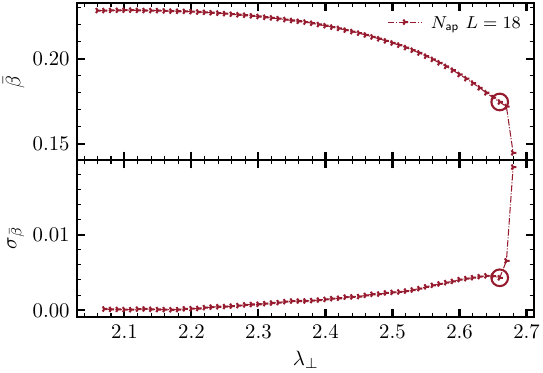}%
            }

            \subfloat[\label{subfig:alpha_fit}]{%
            \includegraphics[width=0.99\columnwidth]{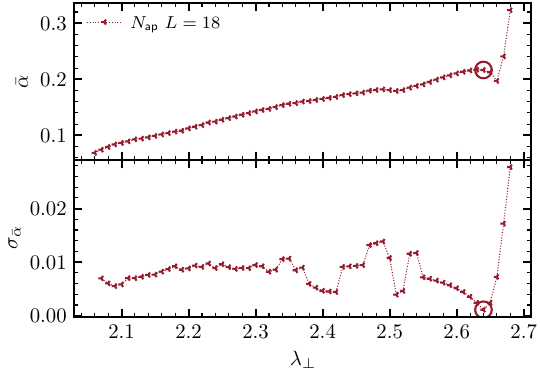}%
            }
            \caption{
                The two figures show a representative result for the exponents of the square lattice Heisenberg bilayer received by the described procedure for $L=18$. 
                \Cref{subfig:beta_fit} shows a result for the exponent $\beta$ and \Cref{subfig:alpha_fit} shows a result for the exponent $\alpha$. 
                The averages of the exponents over the considered couplings are plotted in the upper panels and the corresponding standard deviations in the lower panels as function of the coupling.
                The circles mark the local minimum in the standard deviation and the associated exponent which is extracted for the given system size.
                }
            \label{fig:fit_result}
        \end{figure}

        Then we start with the last determined value on the grid which is closest to the divergence of the flow and take it as the upper bound of the fitting interval of the first run. 
        We choose a lower bound so that $n=n_{\mathrm{min}}$ grid points are in the interval. (For the analysis of the melting of the columnar phase the role of upper and lower bound need to be swapped.)
        In this interval a least-square fit of a power law to the values of the observable is performed.
        The fit parameters are restricted to physically reasonable ranges to make them numerically robust.
		Firstly, the $\lambda_{\rm c}$ of the fit may not exceed $\lambda_{\rm c}$ found by the divergence of the flow by more than $\SI{10}{\percent}$.
		Secondly, the exponent is restricted to be between zero and one.
        Thirdly, the prefactor of the power law is set to be positive for $\beta$ and negative for $\alpha$.
		The fitting procedure is repeated for $n+1$ values, i.e., for the lower bound reduced by one step, and iterated up to $n_{\mathrm{max}}$ values in the interval. 
        This yields a set of values for the exponent under study.
        From this set the average and the standard deviation are calculated and assigned to the value of the upper bound of this first run.

        Then the upper bound value is reduced by $\Delta$ and a second run is carried out analogous to the first one.
        Average and standard deviation are assigned to its upper bound.
        Such run can be continued till most of the values of the grid were chosen as upper bound. 
        The averages and standard deviations are plotted as function of the upper bounds in \Cref{fig:fit_result} for a representative analysis.
        Finally, we choose the average exponent of the upper bound where the standard deviation displays a local minimum not too far away from the critical point. 
        Concretely, we restrict the region where to look for the local minimum to at maximum $\approx\SI{10}{\percent}$ below the value of $\lambda_{\rm c}$ from the flow divergence.
        In our analyses, the values $n_{\mathrm{min}}=5$ and $n_{\mathrm{max}}=20$ produced the most consistent results.

    \end{appendix} 
\end{document}